\documentclass{optica-article}

\journal{opticajournal} 

\articletype{Research Article}

\usepackage{lineno}
\usepackage{color,soul}

\begin{document}

\title{Experimental direct quantum communication with squeezed states}

\author{Iris Paparelle,\authormark{1,*} Faezeh Mousavi,\authormark{2,3} Francesco Scazza, \authormark{2,3} Angelo Bassi, \authormark{2} Matteo Paris \authormark{4} and Alessandro Zavatta \authormark{3,5,6,7}}

\address{\authormark{1}Laboratoire Kastler Brossel, Sorbonne Université, CNRS, ENS-PSL Research University, Collège de France, 4 Place Jussieu, F-75252 Paris, France\\
\authormark{2}Dipartimento di Fisica, Università degli Studi di Trieste, Via Alfonso Valerio 2, 34127 Trieste TS, Italy\\
\authormark{3}Istituto Nazionale di Ottica, Consiglio Nazionale delle Ricerche, Edificio Q2 Area Science Park Strada Statale 14, km 163,5, 34149, Basovizza TS, Italy\\
\authormark{4} Quantum Technology Lab, Università degli Studi di Milano, I-20133 Milano, Italy\\
\authormark{5}QTI s.r.l., Largo Enrico Fermi, 6, 50125 Firenze FI, Italy\\
\authormark{6}Istituto Nazionale di Ottica, Consiglio Nazionale delle Ricerche, Largo Enrico Fermi, 6, 50125 Firenze FI, Italy\\
\authormark{7}European Laboratory for Non-linear Spectroscopy (LENS), University of Florence, Via Nello Carrara, 1, 50019 Sesto Fiorentino FI, Italy}

\email{\authormark{*}iris.paparelle@lkb.upmc.fr} 


\begin{abstract*} 
Quantum secure direct communication (QSDC) is an evolving quantum communication framework based on transmitting secure information directly through a quantum channel, without relying on key-based encryption such as in quantum key distribution (QKD). Optical QSDC protocols, utilizing discrete and continuous variable encodings, show great promise for future technological applications. We present the first table-top continuous-variable QSDC proof of principle, analyzing its implementation and comparing the use of coherent against squeezed light sources. A simple beam-splitter attack is analyzed by using Wyner wiretap channel theory. Our study illustrates the advantage of squeezed states over coherent ones for enhanced security and reliable communication in lossy and noisy channels. Our practical implementation, utilizing mature telecom components, could foster secure quantum metropolitan networks compatible with advanced multiplexing systems.

\end{abstract*}

\section{Introduction}
In response to the fundamental concern of security in information and communication systems, quantum communication (QC) offers a solution based on quantum mechanical laws, e.g., the no-cloning theorem, the uncertainty principle, and Bell’s theorem\cite{pirandola2020advances,zawadzki2021advances}. While quantum key distribution (QKD) is an advanced information-security solution for private key negotiation between legitimate parties over the quantum channel, quantum secure direct communication (QSDC) is another advantageous QC approach, which was proposed in 2002\cite{long2002theoretically} to allow the direct transmission of secret messages without setting up a private-key session beforehand\cite{zawadzki2021advances}.
QSDC protocols are mainly categorized into entanglement-based (i.e., two-step \cite{deng2003two}) and single photon–based (such as the DL04 \cite{deng2004secure}) schemes. Also, a one-step QSDC approach based on one round of signal distribution via hyperentanglement has been proposed\cite{sheng2022one}. Moreover, measurement-device-independent protocols for QSDC have been developed, which can eliminate the loopholes associated with imperfect measurement devices\cite{zhou2020measurement, ying2022measurement}. Even more, device-independent QSDC protocols have been recently proposed to remove all possible attacks related to imperfect devices\cite{zhou2020device, zhou2022one, zhou2023device}. The security proof of QSDC protocols has been presented for a two-way protocol in noisy and lossy channels\cite{hu2020security}. Additionally, the security of DL04 and entanglement-based QSDC protocols has been analyzed using Wyner's wiretap channel theory, and their secrecy capacity has been calculated\cite{qi2019implementation,wu2019security}. More recently, in conditions compatible with real-world experimental settings, the security of QSDC has also been numerically analyzed by considering all practical imperfections, such as detector efficiency mismatch, side-channel effect, and source imperfections \cite{ye2021generic}.

Hitherto, QC schemes have been developed over optical fiber and free-space channels, based on discrete-variable (DV) and continuous-variable (CV) systems. For DV systems, bits are typically encoded on the discrete-valued parameter of photons (e.g.~their polarization), and single-photon detectors are employed. 
On the other hand, in CV approaches which were first proposed by Ralph\cite{ralph1999continuous}, information is encoded (as a Gaussian modulation or a discrete alphabet) into the quadratures $X_1$ and $X_2$ of quantized electromagnetic fields, which can be accessed through shot-noise-limited homodyne or heterodyne detection techniques. As their variance is constrained by the uncertainty relation $\Delta X_1 \Delta X_2 \geq 1/4$, with coherent states being minimum uncertainty states (i.e. $\Delta X_1 = \Delta X_2$ = 1/2), $X_1$ and $X_2$ cannot be simultaneously measured with full accuracy for any given quantum state. Thus, the eavesdropper can't read both quadratures without degrading the state. 
CV methods benefit from their independence of single photon sources and detectors, and their implementation with off-the-shelf components working at room temperature. This affords exciting possibilities to develop robust metropolitan quantum networks built upon  mature technologies inherited from classical communications, hence promising low-cost implementations \cite{pirandola2020advances, diamanti2015distributing, chen2023continuous}. Specifically, CV-QKD protocols have matured over the last two decades from proof-of-concept laboratory experiments\cite{grosshans2003quantum} to different stages of development towards real-life implementations such as in-field demonstrations\cite{fossier2009field}, network integrations\cite{karinou2018integration,chen2023continuous}, and very long distance connections\cite{zhang2020long}.  

An important class of quantum optical resources that have been proposed for implementing CV-QC protocols with higher security (than coherent states) are squeezed states, in which the noise is reduced below the (vacuum) shot-noise along one of two orthogonal field quadratures (while being greater than the shot-noise along the other quadrature, according to the uncertainty relation). This anti-noise property thereby increases the precision of measurements along its orthogonal direction. The reduction in noise 
is particularly advantageous for the legitimate receiver aware of the squeezing direction\cite{ralph2000security,hillery2000quantum,cerf2001quantum,gottesman2003secure}. Theoretically, squeezed-state CV configurations can be more tolerant in purely lossy channels and enable enhanced robustness against highly noisy ones \cite{garcia2009continuous,usenko2011squeezed}. These benefits were experimentally demonstrated\cite{madsen2012continuous,jacobsen2018complete}, and composable security was achieved for combining different cryptographic applications in a unified and systematic way\cite{gehring2015implementation,jain2022practical}. Furthermore, the advantages of CV protocols using squeezed states against different kinds of attacks (individual, collective, and coherent) have been discussed in more detail in Refs. \cite{hillery2000quantum, cerf2001quantum, garcia2009continuous, usenko2011squeezed, madsen2012continuous,jacobsen2018complete, gehring2015implementation,jain2022practical}.

QSDC has been successfully realized heretofore 
in DV encodings based on the DL04 \cite{qi2019implementation,hu2016experimental,pan2020experimental,zhang2022realization,liu2022fiber} and EPR\cite{zhu2017experimental,zhang2017quantum} protocols in both free-space and optical fiber channels. Except for the latter experimental realization\cite{zhang2017quantum}, where quantum memories based on atomic ensembles have been exploited, other works circumvented the quantum memory requirement by delaying the photonic qubits using fiber coils. Recently, with an outlook to large-scale quantum communication networks, the feasibility of a 15-user QSDC network has also been demonstrated\cite{qi2021user}. 
Motivated by the low cost, excellent integrability with existing optical communication systems, and easy implementation from state preparation to measurement, coherent state CV-QSDC has  been proposed as a complementary approach and its security against various attacks has been proved\cite{srikara2020continuous}. To exploit the enhanced noise properties of quadrature-squeezed light, CV-QSDC protocols based on squeezed states have also been put forward, allowing for extracting the weak signals in highly lossy channels\cite{srikara2020continuous, yuan2015continuous,yu2016novel,chai2019novel,cao2021continuous}. Recently, an experimental work, which realized a CV-QSDC protocol based on Gaussian mapping, proposed a parameter estimation scheme for signal classification under the actual channels \cite{cao2023realization}. Nevertheless, none of the CV-QSDC protocols employing squeezed quantum states have been implemented to date.   

In this work, we propose a novel practical CV-QSDC scheme to investigate how and if squeezed light can outperform when compared with coherent states. We realized a proof-of-principle experiment entirely based on optical fibers using homodyne detection. Analyzing the two cases where the transmitter (Alice) or receiver (Bob) possesses a squeezed state source, we analytically and numerically study the two versions of the protocol, symmetric and asymmetric, by considering beam splitter attacks. Furthermore, we study the effect of squeezing level and consider two scenarios regarding the relative phase of coherent states, as phase-locked and randomized. Due to practical implementation issues, we choose the asymmetric protocol for the experimental verification, applying amplitude squeezing, i.e., squeezing along the quadrature direction perpendicular to the phase of the (phase-locked) coherent states. In addition to this proof-of-principle demonstration of CV-QSDC protocol, we reveal the benefit of squeezed states over coherent ones for achieving higher secrecy against a beam splitter attack and better performance in lossy channels. 

\section{Results}
Building upon the key ideas of Ref. \cite{deng2003two} and Ref. \cite{srikara2020continuous}, we propose new schemes for CV-QSDC, which allow for practical realizations in a table-top experimental setup. In the following, we present the CV-QSDC protocols, as well as an analysis of their numerical and experimental performances.
\subsection{The Protocols}
In this section, we illustrate the protocols: We describe modified versions of the protocol proposed by Ref. \cite{srikara2020continuous} to increase its experimental feasibility and performance. 
Two distinct schemes are examined and labeled henceforth as ``symmetric'' and ``asymmetric''. In both versions, the quantum channel used by the sender and the receiver is memoryless. This legitimate channel can be used in both directions (experimentally, this can be implemented e.g. with circulators). Another memoryless channel will also be considered: the wiretap channel. This represents the channel between the legitimate users and the eavesdropper. 
\subsubsection{Symmetric protocol}

\begin{figure}[ht!]
\centering\includegraphics[width=0.9\linewidth]{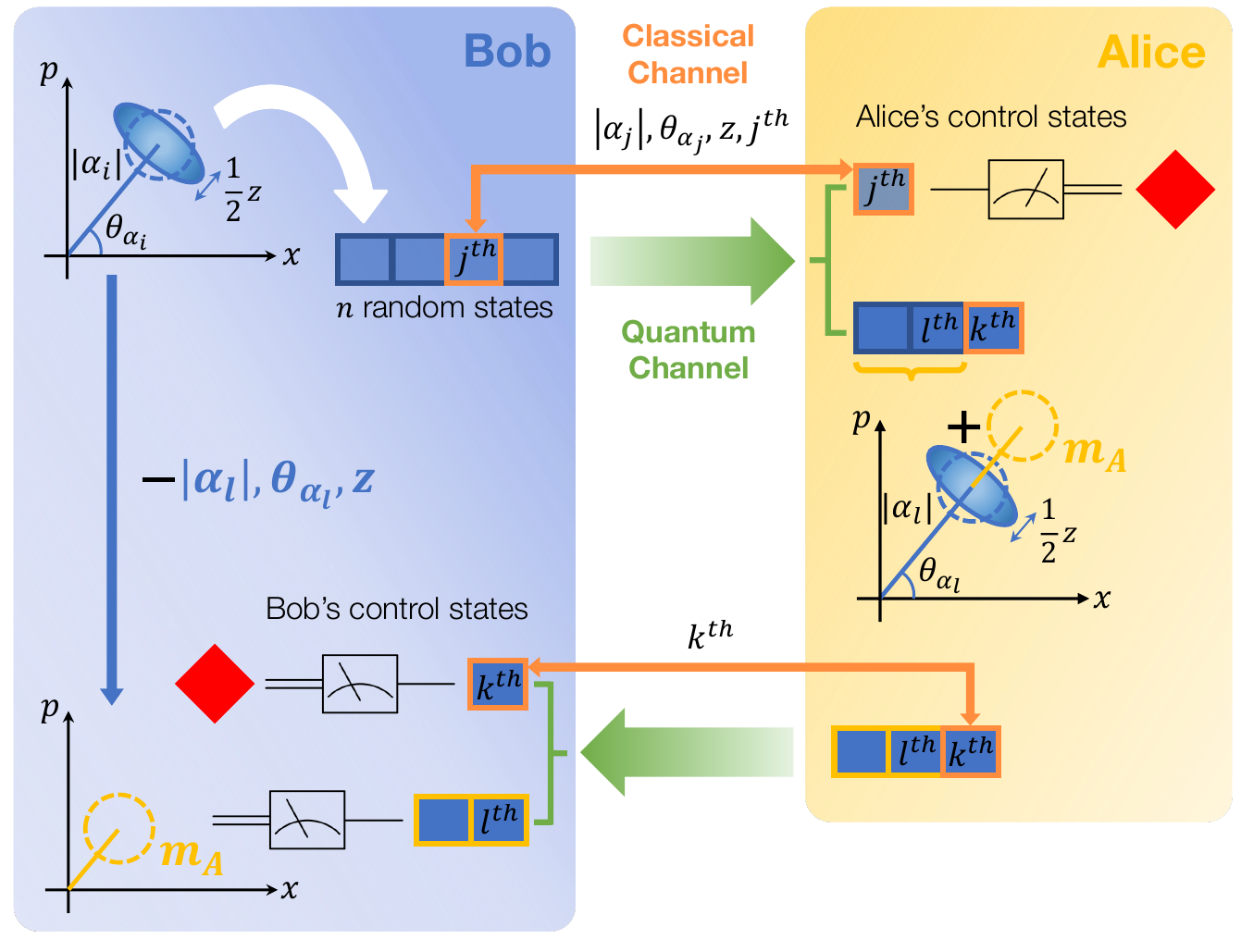}
\caption{Illustration of the proposed symmetric protocol inspired by Ref. \cite{srikara2020continuous}. Alice and Bob send quantum states through a quantum channel (green arrows) and exchange classical information through a classical channel (orange arrows). Bob sends random squeezed states to Alice, they check for an eavesdropper with control states chosen by Alice and exchanging classical communication. If the results are compatible, Alice encodes a message by displacement (or attenuation), leaving some untouched decoy states, and Bob then reads the message by homodyne detection subtracting the initial amplitudes.}
\label{fig:fig1}
\end{figure}

\begin{figure}[ht!]
    \centering
    \includegraphics[width=0.85\linewidth]{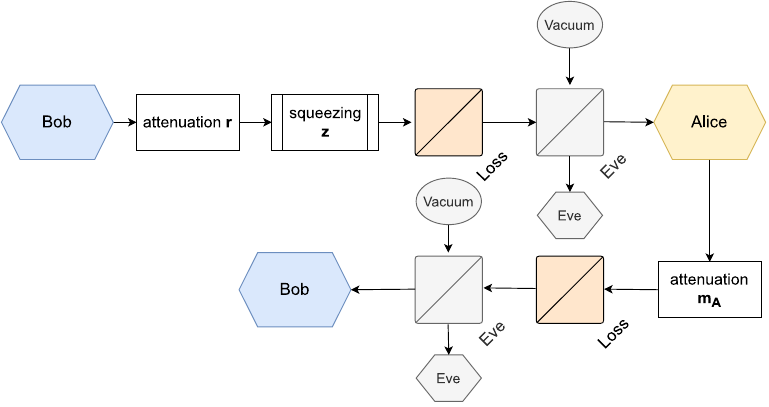}
    
    \vspace*{15pt}
    \includegraphics[width=0.7\linewidth]{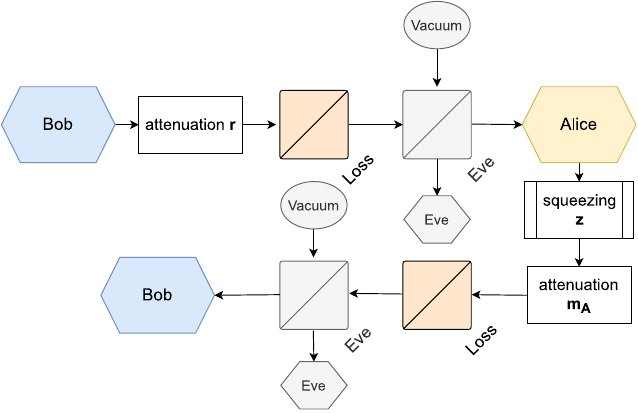}
    \caption{Illustration of the symmetric (top) and asymmetric (bottom) protocols. In the symmetric protocol, Bob (blue) initializes his quantum states by applying both attenuation and squeezing, then sends them to Alice (yellow) via a quantum channel, which is subject to losses (orange beam splitter) and to a beam-splitter attack by Eve (grey). Alice encodes her message via attenuation and uses the channel back to communicate to Bob, again subject to losses and eavesdropping. In the asymmetric case (bottom), Bob only initializes the states via attenuation and Alice applies squeezing. }
    \label{fig:fig2}
\end{figure}

This scheme is illustrated in Figure \ref{fig:fig1}, the label ``symmetric'' refers to the case where squeezing is applied in the initialization by Bob and the feature travels in both directions of the channel, from Bob to Alice and from Alice to Bob. In the ``asymmetric'' version elaborated in this paper, the squeezing will instead be traveling from Alice to Bob in one direction only.  The difference between the two protocols is shown in Figure \ref{fig:fig2}. 
The symmetric protocol proceeds along the following steps: 
\begin{enumerate}

    \item Bob, a legitimate information receiver, prepares a sequence of $n$ coherent states $|\alpha_i\rangle$ with amplitude and phase $|\alpha_i|$ and $\theta_{\alpha_i}$ $(i \in \{1, \ldots ,n\} = I)$, respectively, both chosen randomly from uniform distributions to maximize unpredictability. He then applies squeezing to those states so that the uncertainty of the squeezed state $S(z)|\alpha_i\rangle$ is below the vacuum level when measured along $\hat{x}_{\theta_{\alpha_i}} = \cos(\theta_{\alpha_i}) \hat{x} + \sin(\theta_{\alpha_i}) \hat{p}$. We employ a fixed amount of squeezing $z$ with a squeezing direction perpendicular to the direction of the coherent states, to maximize the precision of the subsequent measurements, while in Ref. \cite{srikara2020continuous} squeezing was randomized.
    
    \item Bob sends the prepared $n$ states $S(z)|\alpha_i\rangle$ $(i \in I)$ to Alice via a quantum channel.
    
    \item Alice uses an optical switch to randomly select a subset $J \subset I$ of the incoming states as control states and measures them via homodyne detection. She sends the indexes $\{j \in J\}$ to Bob via a classical communication channel, whereas she injects the message states (the remaining states) into an optical delay or a quantum memory.
    
    \item Bob shares the information $|\alpha_j|$ and $\theta_{\alpha_j}$ $(j \in J)$ for the control states via the classical channel with Alice, who verifies the correspondence between her results and this data to evaluate the losses in the quantum channel, thus checking for eavesdropping. If the measured values correspond to a tolerable limit, then they continue; otherwise, she discards the protocol, and they start again.
    
    \item Alice chooses another subset $L \subset I\setminus J$ of states from the remaining states and encodes her message $m_A \in \mathbb{R}$ on them by applying an attenuation of $\sqrt{m_A}$. In Ref. \cite{srikara2020continuous}, a displacement operation was used instead. However, implementing displacement is experimentally costly, as another laser is required as well as very precise phase control over the original state and the one to be added. The remaining states $K = ((I\setminus J)\setminus L)$ are used as decoy states. The decoy states allow checking for statistics influenced by channel properties (such as Eve), that would, e.g., reduce the squeezing amplitude. Moreover, they are useful to disorient Eve about the message.
    
    \item Alice sends all the $I\setminus J$ states to Bob via the same quantum channel.
    
    \item Bob performs homodyne measurement on the incoming states along the squeezed quadrature direction, thus optimizing his uncertainty, and he checks for eavesdropping by comparing to the corresponding $|\alpha_k|$ $(k \in K)$ values for the decoy states. For this, Alice communicates their indices via a classical communication channel. Bob also verifies that the uncertainty on measurements is compatible with the initially applied squeezing. In fact, the statistics are the same for all states as they were all submitted to the same amplitude squeezing.
    
    \item Bob can retrieve the message in the other states by dividing the amplitude of the message states by the amplitude $|\alpha_i|$ of the initial states and thus obtaining the message encoded in the corresponding retrieved attenuation values.
\end{enumerate}

\subsubsection{Asymmetric protocol}

\begin{figure}[b!]
    \centering
    \includegraphics[width=0.9\linewidth]{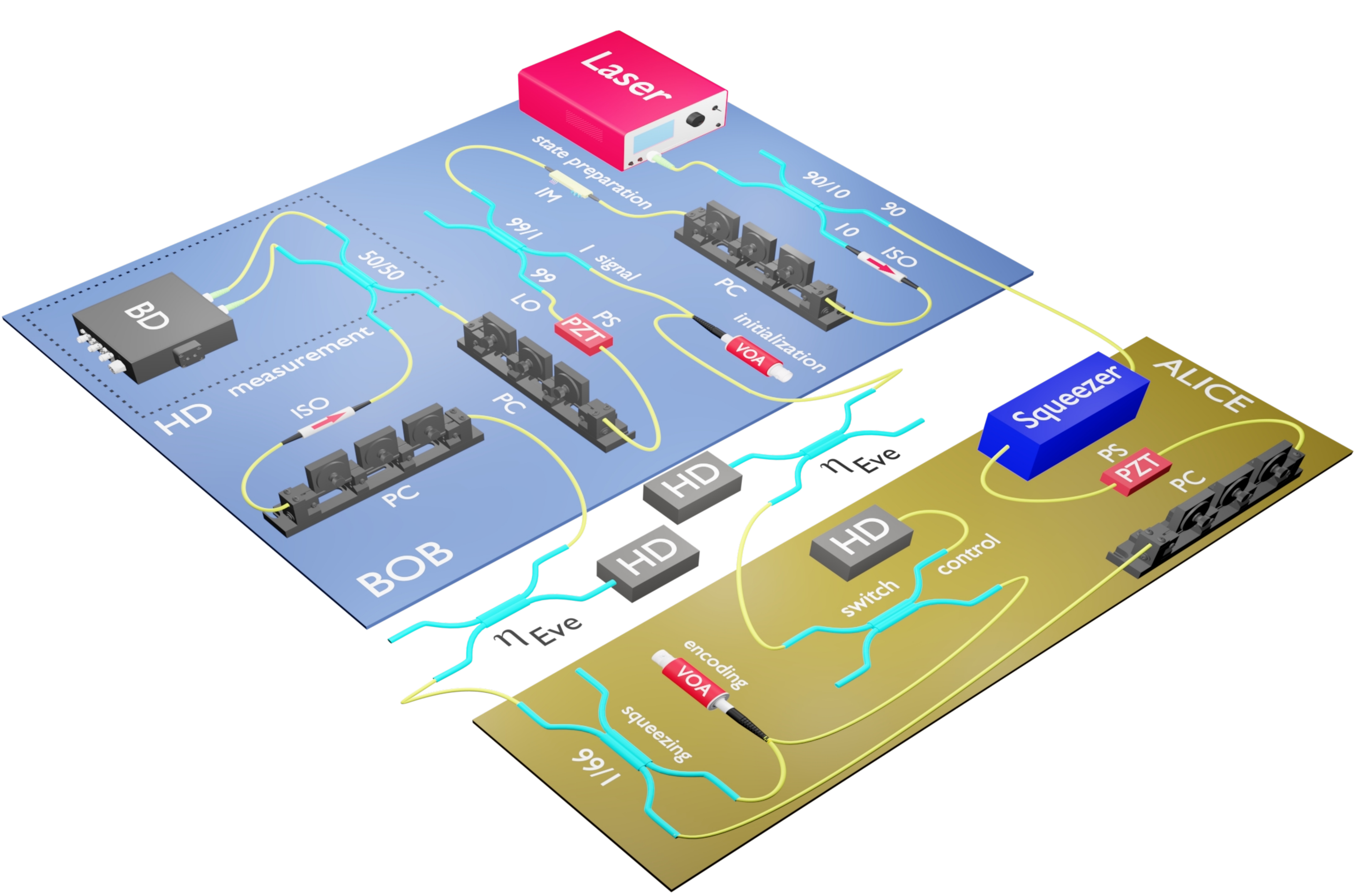}
    \vspace*{5pt}
    \caption{Experimental setup for implementing the asymmetric protocol in case of a beam splitter attack. In this setup, Bob and Alice are indicated by blue and yellow background rectangles, respectively, while Eve1 and Eve2 (she intervenes twice, see cases (a) and (c) of Fig. \ref{fig:fig7}(top)) are shown as beam-splitters In this scenario, Bob sees Eve’s interventions as losses, she acts with two beam splitters of transmissivity $\eta_E$ (the first one can also be modeled with a variable optical attenuator (VOA)). The measurement system includes the polarization controller (PC), isolator (ISO), and homodyne detector (HD), which itself consists of a 50/50 beam splitter and balanced detector (BD). The initialization is applied by Bob through a VOA. Before that, he generates the pulses with an intensity modulator (IM), after propagating the continuous waves of the laser through ISO and PC. 99\% of these pulses go to the local oscillator branch, which together with a piezoelectric (PZT) phase shifter (PS) and a PC, generates the phase control circuit. To generate the squeezed states by Alice, 90\% of the continuous waves are propagated through the squeezer, a PZT PS, and PC, before coupling to the signals via a 99/1 beam splitter. Alice measures some of the states Bob sent, as control states, with a switch.    }
    \label{fig:fig3}
\end{figure}

In this new version of the protocol, coined ``asymmetric'', Alice applies the squeezing to the message states after encoding the message, instead of Bob applying it to all the states and at the beginning of the protocol. In this way, the squeezing feature travels only in one direction in the legitimate channel. As Alice does not know the $\theta_{a_i}$ (the phases of coherent states sent by Bob), her squeezing will in general not be along the perpendicular direction. However:

\begin{itemize}
    \item Alice or Bob can agree on a fixed direction (for example, $\theta_{\alpha_i} = 0$ $\forall i$ and $S(z)$ along $\hat{x}$), which is also easier to implement, not requiring fast phase control. This case was implemented experimentally as shown in Figure \ref{fig:fig3} and Figure \ref{fig:fig7} (in the Materials and Methods section).
    
    \item Alice can communicate the squeezing direction to Bob after the states have traveled the quantum channel and they checked for eavesdropping, and Bob applies the protocol of Ref.~\cite{srikara2020continuous}. 
\end{itemize}

As explained in the Discussion section, the first solution for this asymmetric protocol is experimentally more convenient to be implemented, and  security is not compromised. In the following subsections, we present a security analysis for the case of a beam splitter attack: Eve uses a beam splitter to intercept states in both uses of the quantum channel. Other types of attacks will be considered in the Discussion section. If Alice and Bob use the same quantum channel in two different directions, Eve uses one beam splitter in the middle of the channel. If they use two different channels, we consider Eve to be bound to use two different beam splitters with the same optimal transmissivity. 

\subsection{Numerical security analysis}
\begin{figure}[b!]
    \centering
    \includegraphics[width=\textwidth]{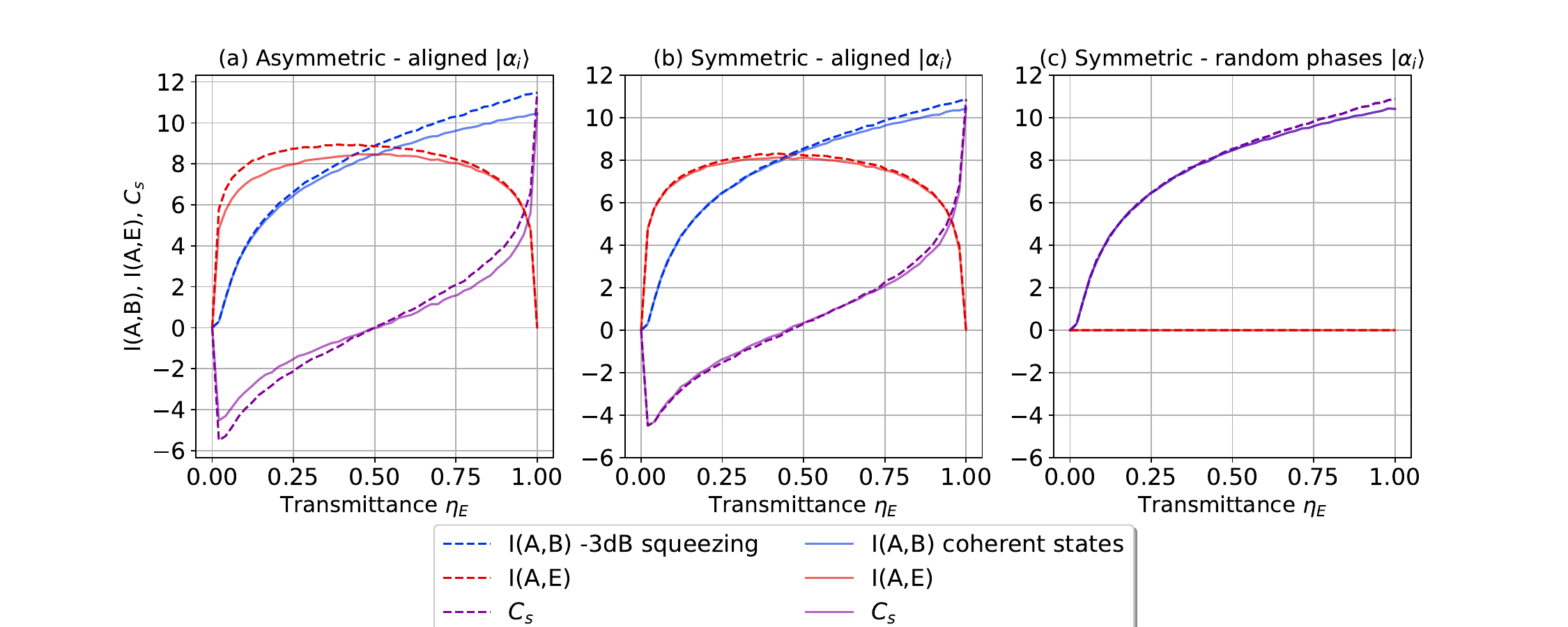}
    \caption{Comparison between different protocol implementations for varying $\eta_E$. (a) Alice applies squeezing to coherent states at fixed phase, (b) Bob applies squeezing to coherent states, again without randomness in their phase, (c) Bob applies the squeezing and uses random coherent states’ phase. Solid lines represent the case with only coherent states, while dotted lines refer to applying a squeezing of –3 dB. Red curves represent the mutual information between Alice and Eve, blue curves represent the mutual information between Alice and Bob, and purple curves indicate the secrecy capacity of the protocol. In the case of a constant coherent state’s phase, the squeezing effect is more effective when it is applied by Alice. The case of the random coherent state phase gives null information to Eve but is more difficult to be implemented. }
    \label{fig:fig4}
\end{figure}

In our numerical simulations we consider three different cases: Alice applies squeezing to coherent states all aligned with one another (see the first case of section 2.1.2, asymmetric protocol); Bob applies squeezing to coherent states all aligned with one another; Bob applies squeezing to coherent states with random phases (see section 2.1.1, symmetric protocol). 
Our security analysis is then based on the calculation of the secrecy capacity. In fact, according to Wyner’s wiretap channel theory\cite{wyner1975wire}, a positive secrecy capacity $C_s$ implies that reliable transmission at rates up to $C_s$ is possible in approximately perfect secrecy. Effectively achieving this maximum rate, however, requires additional strategies, such as optimizing the encoding, tailoring the input distribution to the channel when feasible, improving the decoding process, and employing appropriate error-correction techniques. The following methods have been used in the literature: (de)encoding based on quantum error correction\cite{pan2023free}, secure coding based on the universal hashing families\cite{qi2019implementation}, frequency coding schemes\cite{hu2016experimental}, and privacy amplification in Gaussian-mapping based protocols\cite{cao2021continuous}.

In our analysis, we focus on the calculation of the secrecy capacity as an upper bound on achievable communication rates and as a comparative metric for evaluating the potential performances of the different implementations.

The secrecy capacity is defined as:
\begin{equation}
C_s = \max_{p_A} [I(A,B) - I(A,E)]
\end{equation}
where $I(A,B)$ and $I(A,E)$ are the mutual information between Alice and Bob and between Alice and Eve, respectively. Moreover, in the calculation of $I(A,E)$ we ought to consider the best conditions for Eve, that is to say, the worst-case scenario for Alice and Bob. 
The Gaussian formalism (see Methods) allows the representation of Gaussian states, which include coherent states and squeezed states, Gaussian operations -- such as the application of squeezing, the effect of beam splitters and displacement -- and measurements including homodyne detection\cite{serafini2023quantum}. In this way, experimental results can be predicted, as all the elements of the experimental setup fall within the set of Gaussian operations. More details about this formalism are reported in the Materials and Methods section, while for detailed calculations please refer to the Supplementary Material.   

Concerning the cases in which the phases are all equal to zero, analytical calculations suffice and are confirmed by numerical simulations, which means that the asymptotic formulas correspond to the case with a finite number of states.

\bigskip
\textbf{Alice implements the squeezing:}
\begin{align}
    I_{\text{asym}}(A,B) &= \log_2\left(\frac{0.01 \eta_E^2 \eta_L^2 \text{Var}(x\sqrt{m_A})}{\frac{1}{2}\left(1-0.99\eta_E \eta_L (1-z^2)\right)}\right) \label{eq:asymAB} \\
    I_{\text{asym}}(A,E) &= \log_2\left(\frac{0.01 \eta_E (1-\eta_E) \eta_L^2 \text{Var}(x\sqrt{m_A})}{\frac{1}{2}\left(1-0.99(1-\eta_E)\eta_L (1-z^2)\right)}\right) \label{eq:asymAE} \\
    C_{\text{s asym}} &= I_{\text{asym}}(A,B)-I_{\text{asym}}(A,E) \label{eq:asymC}
\end{align}

\textbf{Bob implements the squeezing:}
\begin{align}
    I_{\text{sym}}(A,B) &= \log_2\left(\frac{0.01 \eta_E^2 \eta_L^2 \text{Var}(x\sqrt{m_A})}{\frac{1}{2}\left(1-0.99\eta_E^2 \eta_L^2 \text{Var}(m_A)(1-z^2)\right)}\right) \label{eq:symAB} \\
    I_{\text{sym}}(A,E) &= \log_2\left(\frac{0.01 \eta_E^2 \eta_L^2 \text{Var}(x\sqrt{m_A})}{\frac{1}{2}\left(1-0.99\eta_E (1-\eta_E) \eta_L^2 \text{Var}(m_A)(1-z^2)\right)}\right) \label{eq:symAE} \\
    C_{\text{s sym}} &= I_{\text{sym}}(A,B)-I_{\text{sym}}(A,E) \label{eq:symC}
\end{align}

In these equations, $\eta_E$ is the transmissivity of Eve’s beam splitter, $\eta_L$ indicates the transmissivity of the beam splitter representing losses, $x$ is the variable corresponding to the initial quadrature values along the $\hat{x}$ direction (put by Bob, randomly), $m_A$ represents the variable corresponding to the message sent by Alice, and $z$ is the squeezing value: $z=10^{(\text{squeezing(dB)}/10)}$. The values of 0.01 and 0.99 correspond to $\eta_{\text{sq}}$ and $1-\eta_{\text{sq}}$, with $\eta_{\text{sq}}$ being the transmissivity of the beam splitter that mixes the signal with a vacuum squeezed state to get a squeezed signal. All the numerical results are presented in Figure \ref{fig:fig4}: the mutual information between Alice and Bob, the one between Alice and Eve, and the associated secrecy capacity. All quantities are plotted for varying transmissivity $\eta_E$ of Eve’s beam splitters. In the three cases, a squeezing of –3 dB is applied to the coherent states to be compared to the case where only coherent states are used. According to the analytical formulas, Alice and Bob’s mutual information is monotonically increasing with $\eta_E$, which means that more light goes to the legitimate sender and receiver. Additionally, the mutual information between Alice and Eve is null for $\eta_E=0$ and $\eta_E=1$ as Eve cannot retrieve the message if she does not intercept any information in one of the two directions of communication. In fact, when Bob sends the quantum states to Alice, Eve intercepts a fraction of light equal to $\sqrt{1-\eta_E}$ and when Alice sends them back to Bob with the encoded message, Eve then intercepts a fraction $\sqrt{\eta_E} \sqrt{1-\eta_E}$, where $\sqrt{\eta_E}$ is the fraction that has arrived to Alice from Bob. By comparing the plots in Fig.~\ref{fig:fig4}, it can be noticed that for the phase-locked coherent states, the effect of squeezing is stronger when applied by Alice, as the squeezed states are less subjected to losses. Thus, the asymmetric configuration achieves higher secrecy for the channel. Also, the last plot shows that randomness reduces the mutual information between Alice and Eve to nearly zero, since in this simulation Eve has no information on the optimal measurement direction.

\begin{figure}[ht!]
    \centering
    \includegraphics[width=0.9\linewidth]{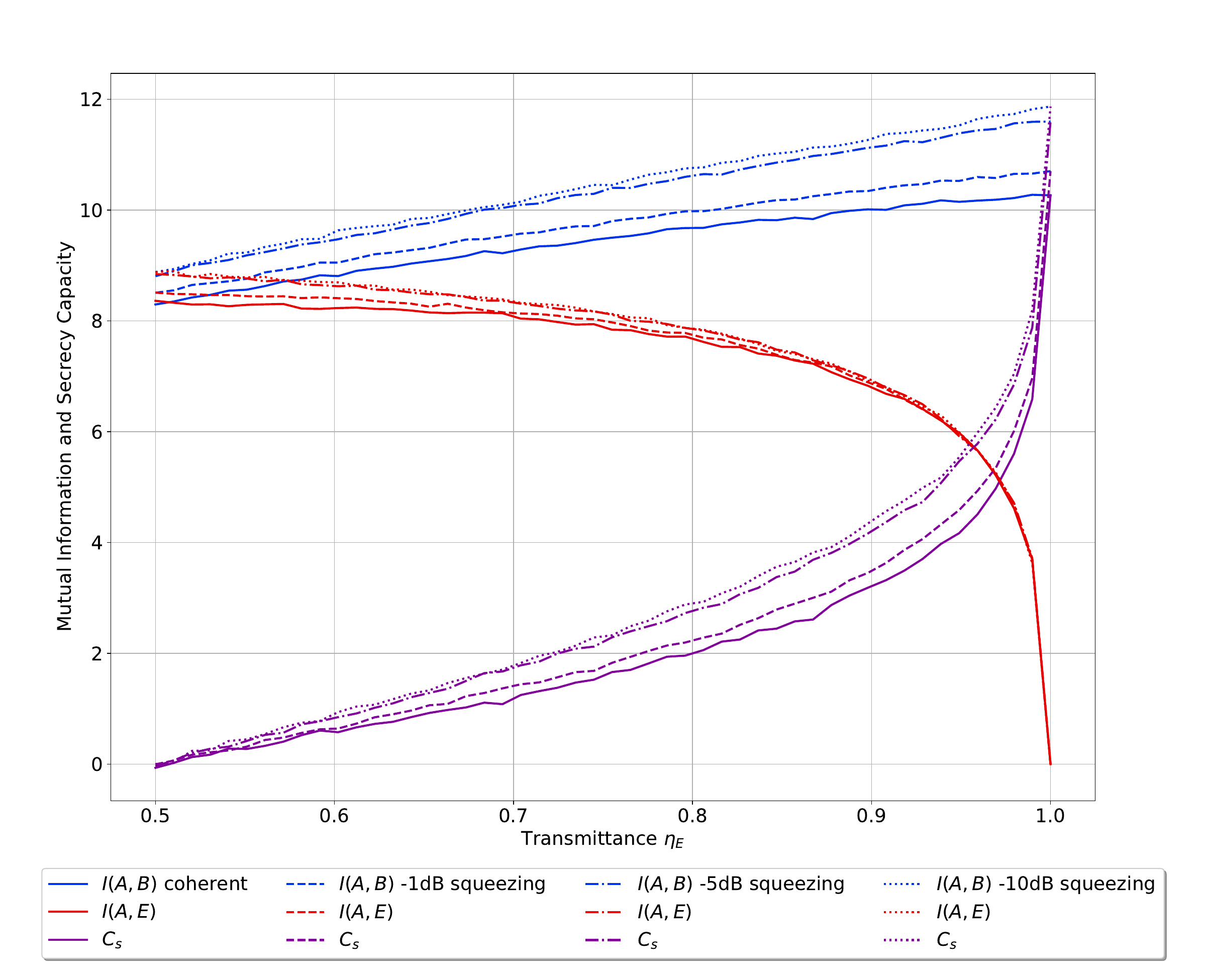}
    \caption{Numerical prediction of the effect of different squeezing levels in the experimentally implemented scheme (i.e., Alice implements squeezing and there is no randomness in the phase of the coherent states used). The solid lines denote the case where only coherent states are used; the dashed, dash-dotted, and dotted lines show results for -1dB, -5dB, and -10dB of squeezing, respectively. We can see a saturation effect reducing the impact of squeezing while approaching -10dB. Red curves denote the mutual information between Alice and Eve, blue curves denote the mutual information between Alice and Bob, and purple curves denote the secrecy capacity of the protocol.}
    \label{fig:fig5}
\end{figure}

Figure \ref{fig:fig5} displays the numerical predictions for different values of squeezing. It is possible to observe a saturation effect, with –5 dB of squeezing being already a nearly optimal value, implying no need to resort to large, current-record squeezing levels of -15 dB. In our proof-of-principle implementation, we were able to implement a squeezing of about –1 dB. 
Analytical formulas as well as numerical simulations show that for both the symmetric and asymmetric protocols, the secrecy capacity is positive for $\eta_E=0.5$, which is true for all squeezing values $z$ as well as all collective losses $\eta_L$ (that act in the same way on Eve and on Bob). In a worst-case scenario, all losses that Bob experiences are intercepted by Eve, this is equivalent to $\eta_L=0$ and Bob experiencing $\eta_E$ of transmission. The protocol is then secure (positive secrecy capacity) only for losses equivalent to $\eta_E=0.5$, i.e. -3dB. With ultra-low loss optical fibers, 3dB are equivalent to 20 km of transmission distance.

\subsection{Experimental results}
We implemented the asymmetric scheme where Alice applies the squeezing to her states, and all states are aligned along the same direction (phase-locked), that is to say, they have the same phase and are equally projected by the homodyne detection along, without loss of generality, $\hat{x}$. The quantum channel and the other elements of the scheme consisted of an optical fiber system with amplitude coding. In fact, in Ref. \cite{srikara2020continuous} it is proposed that both the random initialization of Bob and the message are coded by displacement on coherent states: $|\alpha_i\rangle$ ($\alpha_i\in\mathbb{C}$) and $|m_A (1+i)\rangle$ ($m_A\in\mathbb{R}$). However, by choosing to fix the coherent state’s phase and optimizing the angle between initialization states and message (equal to $0$ degrees), we can restrict ourselves to $|\alpha_i\rangle$ with $\alpha_i\in\mathbb{R}^+$ and $|-m_A\rangle$ with $m_A\in\mathbb{R}^+$, so that we can use attenuators instead of implementing displacement. More details about the experimental setup will be provided in the Materials and Methods section.

\begin{figure}[ht!]
    \centering
    \includegraphics[width=0.9\linewidth]{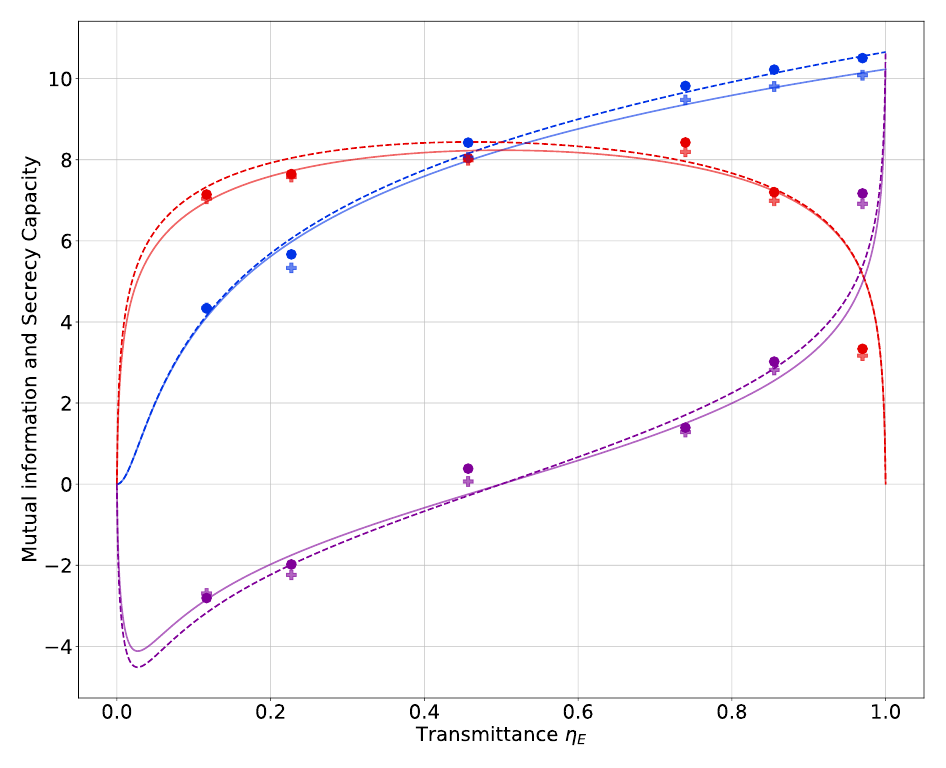}
    \caption{Experimental results and corresponding analytical calculations. The lines represent the analytical calculations: the dotted lines denote the expectations for a squeezing of -1dB, while solid lines denote the case without squeezing (coherent states). The experimental results are denoted by circles in case of -1 dB of squeezing and by crosses in the absence of squeezing. Colors denote different quantities: the mutual information between Alice and Eve (red), the mutual information between Alice and Bob (blue), and the secrecy capacity of the protocol (purple). Error bars are evaluated by error propagation and the boot-strap method, and are smaller than the size of the symbols.}
    \label{fig:fig6}
\end{figure}

The experimental results are presented in Figure \ref{fig:fig6}, compared with the analytical calculations in the same conditions, with –1 dB of squeezing -- the highest level that we were able to inject considering experimental losses. We observe good agreement between calculations and experimental results, and we experimentally reveal the advantage of squeezing especially in the mutual information between Alice and Bob. We also considered the case without squeezing, 
and obtained good agreement between theoretical expectations and experimental results. 
\section{Discussion}
\subsection{Attacks}
To evaluate the secrecy capacity of the protocol and thus its security, the mutual information between Alice and Eve must be calculated explicitly. We thus have to specify which kind of attacks the eavesdropper may use.

The eavesdropper attacks are traditionally divided into three categories: individual, collective, and coherent. In individual attacks, Eve attacks each of the quantum states traveling through the channel individually. She attaches an ancilla system to each of the states, applies the same unitary operation to each composite system, and then measures her part of all the composite systems separately. Collective attacks are more advanced in the sense that Eve still attaches individual ancilla systems to each quantum state, but she measures all the states collectively instead of individually. The most general type, coherent attacks, involves Eve attaching a large ancilla system to all the quantum states sent by Alice. Eve then applies a global unitary operation to the entire composite system, and similar to collective attacks, then performs a global measurement on her part of the system \cite{wolf2021quantum}. Among these attacks, in this work, we theoretically and experimentally focused on the individual attacks (specifically, beam-splitter attack) as a first step towards the security of this proof-of-principle protocol. The beam-splitter attack can be considered the most practical attack for Eve, because it is easily implementable, does not need quantum computers nor quantum memories, and can be mistaken for losses if Alice and Bob do not know their channel sufficiently well.

Another attack strategy on realistic setups of quantum cryptographic systems is the Trojan Horse Attack (THA). The main idea of this attack strategy is not to interact with the photons in transit between Alice and Bob but to probe the devices in Alice’s and Bob’s laboratory by sending some light into them and collecting the back-reflected signal. In this way, Eve, which has a laser and detection system, can acquire knowledge of the properties and functionality of the detectors. The common practical measure to prevent this attack is to add an isolator in Alice’s and Bob’s laboratory to block Eve’s injected pulse, as an ideal isolator would passively stop any THA by a complete extinction of Eve’s pulses. A special kind of THA in QSDC protocols is Invisible Photon Eavesdropping (IPE), in which Eve uses an invisible photon - i.e. a photon to which Alice's and Bob’s detectors are insensitive). So, while Alice and Bob cannot detect Eve’s eavesdropping attack when they check the fidelity of their photons, she can gain full information about the communication. However, Eve’s invisible photons can be filtered out. An optical narrow bandwidth isolator performs this filtering process, which we added before Alice’s and Bob’s detectors in our setup, thereby prohibiting all kinds of THAs in our scheme \cite{jain2014risk,kollmitzer2010applied,cai2006eavesdropping}.

Among collective ones, an effective attack in QSDC protocols is the Teleportation Attack, by which Eve can extract half of the secret bits via intercepting by EPR states and employing the technique of quantum teleportation. This kind of attack will succeed if the security of the quantum channel is examined only after photon transmissions are all finished. However, in our protocol, the error rate is checked after photon transmissions to ensure that the quantum channel is secure before encoding and decoding. Thus, this gap is closed, and this attack is ineffective in our protocol \cite{fei2008teleportation}. 
The effectiveness of collective attacks, such as the teleportation attack, and coherent attacks in our scheme will be the subject of more in-depth investigation in future works.

\subsection{Comparison of the different protocols}
The randomness in the phase of the coherent states that Bob sends has a significant impact on the information that the eavesdropper can get. This is because, even if Eve measures part of the light sent by Bob to Alice, her homodyne detection would not give an indication of the correct phase of the states, as they all have different amplitudes. 
Even for a fixed phase for all the states, Eve could not retrieve the exact phase from measurements; however, her reading of the states would only be multiplied by a factor $\cos(\theta_{\text{Eve}}-\theta)$, $\theta_{\text{Eve}}$ being the phase shift between the local oscillator that Eve is using and the signal, and $\theta$ the one that Bob implements. In the previous calculations, we considered the worst-case scenario, $\theta=\theta_{\text{Eve}}=0$. In Ref. \cite{srikara2020continuous} the squeezing amplitude $z$ was also randomized, but this decreases the mutual information between Alice and Bob, which is monotonically increasing with $z$ and does not decrease the one between Alice and Eve. This is because the squeezing acts only on the precision of the measurement (the variance of the states), not on the mean value; thus, to decrease the reliability of Eve’s measurements, it is better to always use the maximum available squeezing level.

\subsection{Experimental considerations}
The protocol we implement uses off-the-shelf optical elements, it works at telecom wavelengths and does not require single-photon detectors, thus increasing its appeal for future telecommunications technologies. Furthermore, as the communication is direct there are no problems linked to key storage. Moreover, it can exploit the advantage brought by squeezed light. 

The experimental setup is shown in Figure \ref{fig:fig3}. With our resources we implemented separately the case where Alice measures (the control states), the case where Eve measures for the first time (between Bob and Alice), for the second time (between Bob and Alice), and the final case where Bob measures (the message and decoy states). To differentiate these situations (see Figure \ref{fig:fig7} (top)) the balanced detector must be placed after the component representing the last action before the measurement. Experimentally, having Bob implement the squeezing is not convenient as it is destroyed by channel losses and components, such as the attenuators, used by Alice to encode the message. 

In the final experimental setup, Alice and Bob use two different quantum channels. However, circulators could also be used to reduce the number of legitimate channels to one, in this case, Eve should use a single beam splitter to measure the states in the two directions. Concerning the local oscillator, needed by all actors to perform the homodyne detection, a clock could be used by Alice and Eve to create a “local local 
oscillator”\cite{suleiman2022fiber}. This situation supposes that Eve gets all the classical information that Alice and Bob exchange (the clock is part of it) and is equivalent to Bob sending his local oscillator signal to Alice and Eve, as implemented.  
The studied protocols can be used with discrete alphabets, creating bins of attenuation values, but also to transmit continuous signals, for instance, audio signals or sensor readings, directly interfacing sensors, and readout devices. 

Concerning the delay for allowing classical communication, in experimental  2-way direct communication protocols it was already implemented via fiber-based delay lines\cite{hu2016experimental,zhu2017experimental,qi2019implementation,pan2020experimental}, quantum memories\cite{zhang2017quantum}, or quantum-memory-free via secure coding\cite{sun2018design,sun2020toward,zhang2022realization}. Hence, for our future practical works, we will implement our protocol variants based on the mentioned approaches.

\subsection{Conclusions}
In summary, we have studied the feasibility of a practical continuous-variable quantum secret direct communication system, with either coherent or squeezed light sources, based on security analysis using Winer’s wiretap channel theory. According to analytical and numerical calculations for different configurations (including the position of the squeezer, level of squeezing, and relative phase of coherent states), we implemented a practical protocol as an asymmetric scheme with -1 dB squeezing level and phase-locked coherent states. We verified the protocol over an optical fiber channel in the presence of a beam-splitter attack, while achieving higher secrecy in case of a squeezed source over the coherent one. The squeezed source also shows an increased robustness against lossy and noisy channels. The attenuation-based encoding of this protocol makes it more practical than an already proposed scheme based on displacement \cite{srikara2020continuous}. Since this proof-of-principle demonstration of a QSDC protocol relies on CV encoding, which does not require single-photon detectors, it can be rendered compatible with existing mature classical communication systems. By controlling the phase, the measurement of quantum quadrature will be easier than more sophisticated single photon detections. Thus, in case of the availability of low-noise fibers, this keyless QSDC protocol provides an efficient candidate, complementary to QKD, for the development of quantum networks built upon wavelength division multiplexing systems.

\section{Materials and methods}
\subsection{Gaussian formalism}
Light quantization allows us to define quadrature operators $\hat{x}$ and $\hat{p}$ that verify canonical commutation relations. This puts us in the continuous variable description, which would require the adoption of an infinite-dimensional Hilbert space. However, for our security analysis, the restriction to Gaussian states is adopted, using the Gaussian formalism described in Ref. \cite{serafini2023quantum}. This is because our system is coupled linearly with the environment, and the overall Hamiltonian is at most quadratic; in fact, the modeling of quantum dynamics through second-order Hamiltonians is very common for quantum light fields, as the higher terms are inconspicuous and negligible. Gaussian states $\rho_G$ are completely determined by the vector of first moments $\bar{r}$ and the covariance matrix $\sigma$. For a single degree of freedom as in our case:
\[
\bar{r} = (\bar{x}, \bar{p})^T = (\langle \hat{x} \rangle_{\rho_G}, \langle \hat{p} \rangle_{\rho_G})^T; \quad \sigma = \begin{pmatrix} \text{Var}(x) & \text{Cov}(x,p) \\ \text{Cov}(x,p) & \text{Var}(p) \end{pmatrix}
\]
For a coherent state $|\alpha\rangle$ with $\alpha = |\alpha| e^{i\theta}$:
\[
\bar{r} = (\sqrt{2} \text{Re}(\alpha), \sqrt{2} \text{Im}(\alpha))^T; \quad \sigma = \sigma_{\text{vacuum}} = \begin{pmatrix} \frac{1}{2} & 0 \\ 0 & \frac{1}{2} \end{pmatrix}
\]
Gaussian operators are then easily applied to quantum Gaussian states. In this paper, the following operators were used:\\

\textbf{Single-mode squeezing}
\[
\bar{r} \mapsto S(z) \bar{r} \text{ and } \sigma \mapsto S(z) \sigma S(z)^T \text{ with } S(z) = \begin{pmatrix} z & 0 \\ 0 & z^{-1} \end{pmatrix}, \quad z \in [0,1] \text{ in our case}
\]

\textbf{Two-mode beam splitter}
\[
\bar{r} \mapsto S(\theta) \bar{r} \text{ and } \sigma \mapsto S(\theta) \sigma S(\theta)^T
\]
with
\[
S(\theta) = \begin{pmatrix} \begin{pmatrix} \cos\theta & 0 \\ 0 & \cos\theta \end{pmatrix} & \begin{pmatrix} \sin\theta & 0 \\ 0 & \sin\theta \end{pmatrix} \\ \begin{pmatrix} -\sin\theta & 0 \\ 0 & -\sin\theta \end{pmatrix} & \begin{pmatrix} \cos\theta & 0 \\ 0 & \cos\theta \end{pmatrix} \end{pmatrix}
\]
where $\cos^2\theta = \eta$, $\eta$ is the transmissivity of the beam splitter.

For measurement, we consider homodyne detection, which projects the quantum state along a specific direction $\hat{x}_\phi = \cos\phi \hat{x} + \sin\phi \hat{p}$. In fact, the state is mixed with a strong coherent state (local oscillator) at a balanced ($\eta=0.5$) beam splitter, then the detected intensities at the outputs are subtracted, and the results are indeed proportional to the projection on the desired quadrature, $\phi$ depending on the dephasing between the signal to be measured and the local oscillator\cite{serafini2023quantum}.

\subsection{Quantum Information Theoretical Tools}

To quantify the security of the proposed protocols, we utilized elements of classical and quantum information theory. Given two random variables $X$ and $Y$ with supports $\mathcal{X}$ and $\mathcal{Y}$, we define the following quantities:
\begin{itemize}
    \item Entropy: $H(X) = -\sum_{x \in \mathcal{X}} p_X(x) \log p_X(x)$
    \item Conditional entropy: $H(Y|X) = -\sum_{x \in \mathcal{X}, y \in \mathcal{Y}} p_{X,Y}(x,y) \log p_{Y|X}(y|x)$
    \item Mutual information: $I(X;Y) = H(X) - H(X|Y) = H(Y) - H(Y|X)$
    \item Channel capacity: $C = \sup_{p_X(x)} I(X;Y)$
\end{itemize}
The logarithms are in base two to quantify information in bits. These expressions are derived from Refs. \cite{shannon1948mathematical} and \cite{gallager1968information}.
We applied the Wyner wire-tap channel model \cite{wyner1975wire}, assuming transmission over a discrete, memory-less channel subject to wiretapping. This implies the existence of a secrecy capacity $C_s > 0$ such that reliable transmission at rates up to $C_s$ is possible with approximately perfect secrecy. Naming $X$, $Y$, and $Z$ the random variables representing the message, what the receiver gets, and what the eavesdropper understands, according to Wyner, we can write:
\[ C_s = [I(X;Y) - I(X;Z)] \]
For numerical simulations and experimental results, we used the Shannon-Hartley theorem to calculate the mutual information:
\[ I(X;Y) = \log(1 + \text{S/N}) \]
where $S$ is the mean of the received signal and $N$ is the noise. We consider a normalized bandwidth of 1 to calculate the quantities in number of bits per channel use.

\subsection{Numerical Environment}
Analytical calculations were performed to evaluate $C_s$, using the uncertainties of the $\sigma$ matrices of the quantum states for noise $N$, and the mean vector $\bar{r}$ for signal $S$. However, to model a finite number of measurements and randomness, symplectic operations were directly implemented in Python 3.8.3. The signal-to-noise ratio $S/N$ was then replaced by the inverse of the quadratic error in the reading of the message, which was also done for the experimental analysis.

\subsection{Experimental Setup}

\begin{figure}[ht!]
    \centering
    \includegraphics[width=0.5\linewidth]{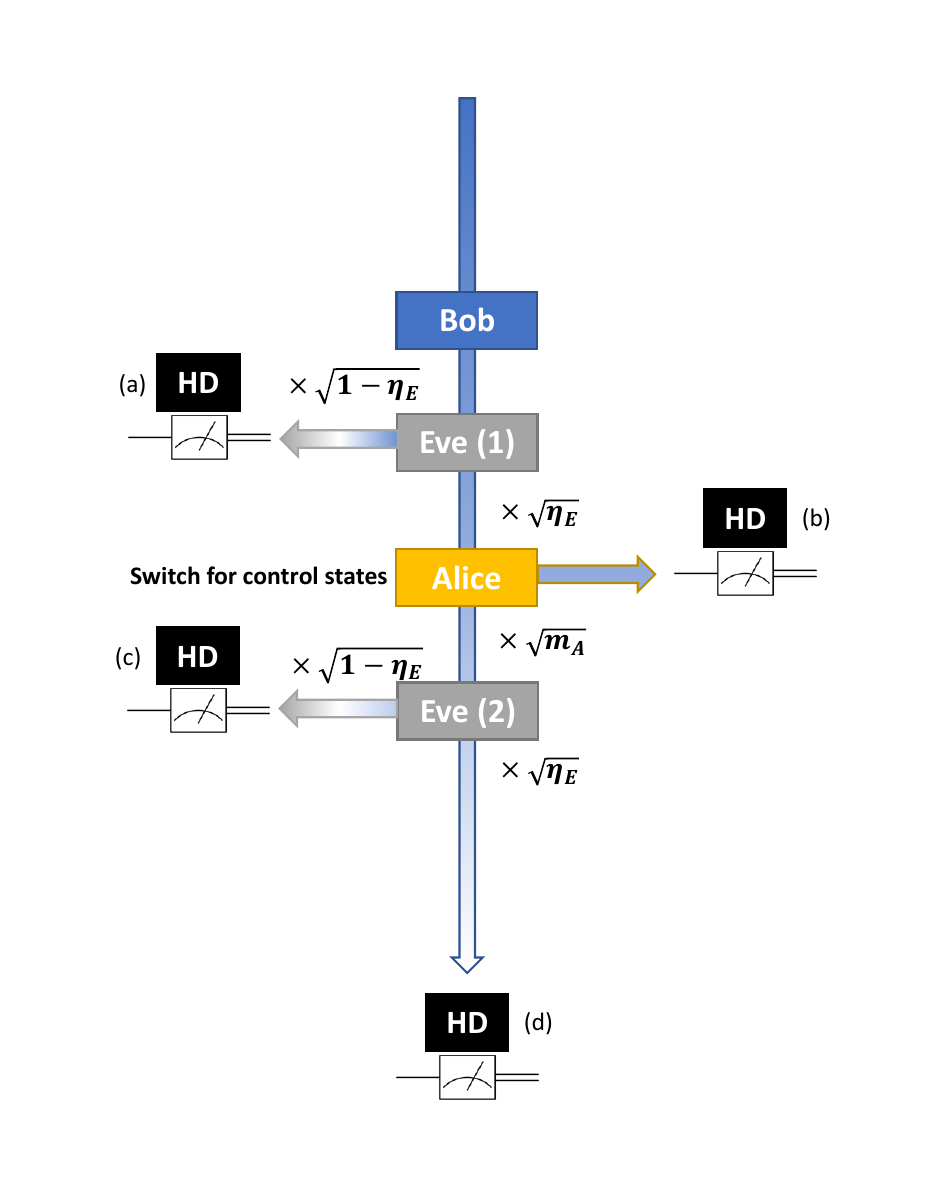}
    \includegraphics[width=\linewidth]{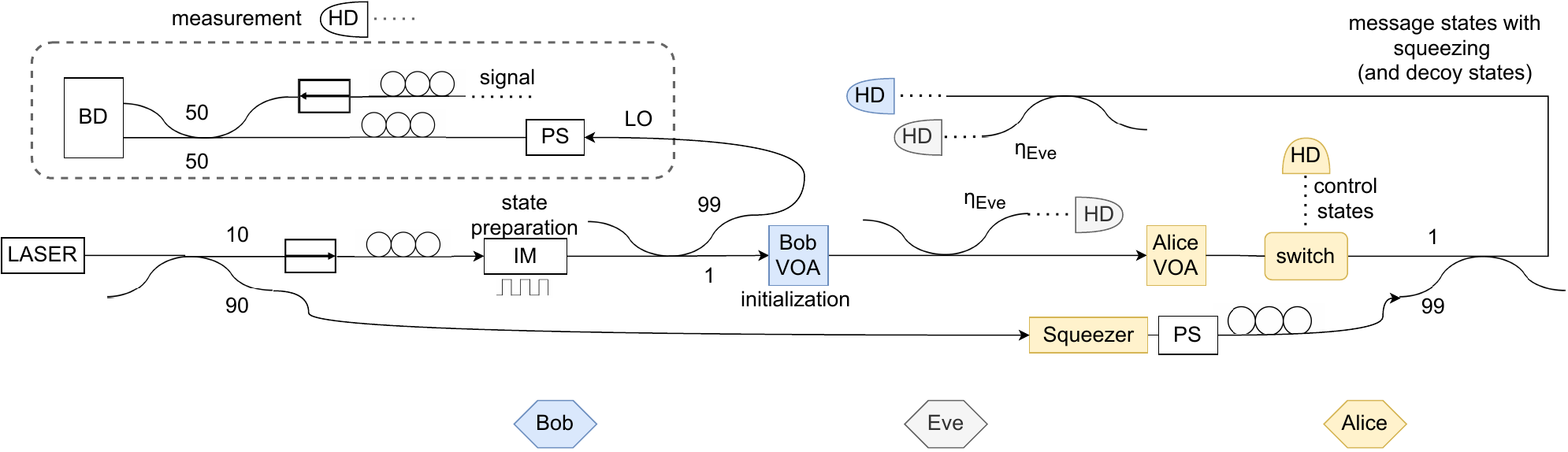}
    \caption{(top) schematic representation of the different measurements done in case of a beam splitter attack: Eve uses a beam splitter between Bob and Alice and measures (a), Alice measures the control states (b), Eve uses again a beam splitter but between Bob and Alice and measures (c), Bob measures the decoy and message states (d). (bottom) Experimental setup to implement the mentioned measurements by Eve, Alice, and Bob. }
    \label{fig:fig7}
\end{figure}

To experimentally investigate the asymmetric protocol in the presence of a beam splitter attack, we followed the indicated diagram in Fig. \ref{fig:fig7} (top) and implemented the setup shown in Fig.\ref{fig:fig7} (bottom) to perform measurements by Eve, Alice, and Bob. In all cases, a CW laser with wavelength $\lambda = 1560$ nm was used to create squeezed light, serving as the local oscillator for the homodyne detections and the signals. The laser output was first split with a beam splitter of transmissivity $\eta = 90\%$, allowing the more powerful branch to pass through a squeezer, which required powers of the order of 500 mW. Squeezed light was then generated using two lithium niobate crystals, resulting in approximately -3 dB of squeezed vacuum \cite{kaiser2016fully}. Alice's experimental application of squeezing differed slightly from the theoretical symplectic operation proposed by Ref. \cite{serafini2023quantum}. In our setup, Alice applied squeezing by mixing her coherent quantum states of the form $\bar{r} = (x,p)^T$, $\sigma = \frac{1}{2} \begin{pmatrix} 1 & 0 \\ 0 & 1 \end{pmatrix}$ with an approximately null squeezed state $\bar{r} = (0,0)^T$, $\sigma = \begin{pmatrix} z^2 & 0 \\ 0 & z^{-2} \end{pmatrix}$ through a beam splitter with transmissivity $\eta_{\text{sq}} = 99\%$, resulting in:
\[ \bar{r} = (0.1x, 0.1p)^T; \quad \sigma = \frac{1}{2} \begin{pmatrix} 0.01 + 0.99z^2 & 0 \\ 0 & 0.01 + 0.99z^{-2} \end{pmatrix}, \quad z \in [0,1] \]

Thus, obtaining a bright state that has a reduced variance along the x direction, although with a reduced amplitude (a factor 1/10, which corresponds to -10dB of losses). 
Concerning the other branch of the 90/10 beam splitter it represents Bob’s laser, with which he generates both his signal (the n quantum initial coherent states) and the local oscillator for homodyne detection. Pulsed coherent states (50 ns) are created and initialized with different amplitudes thanks to a lithium niobate electro-optic intensity modulator (IM) driven by a function generator (FG) and a variable optical attenuator (VOA), respectively.  Concerning the local oscillator, in the experiment all actors have at their disposal the one created by Bob, which would be the worst case. To vary Alice’s attenuations (i.e. her message), we applied different tensions to Variable Optical Attenuators. 
Eve is represented by a VOA (from Bob to Alice) and a beam splitter (from Alice to Bob), this is because we separated the four measurement scenarios, so in order to modulate Eve first effect on the communication, we used directly a VOA. In the second case, a VOA would have destroyed the squeezing as it introduces phase noise and the decoy states would have revealed Eve’s presence. This is because their variance would have been modified (they all follow the same statistics as the squeezing amplitude is always the same).  
Concerning the homodyne detector, it is a commercial model with a bandwidth of 400 MHz, allowing us to use pulses at 10 MHz (50\% duty cycle), and its clearance, for a local oscillator of around 2.5 mW, is approximately 11. The clearance is defined as the ratio between the variance of the homodyne signal measuring the vacuum (but $P_{\text{LO}} = 2.5$ mW) and the electronic noise (variance at $P_{\text{LO}} = 0$ mW).

The efficiency of the photodetectors of the homodyne is 91±0.5\% for both, while the quantum efficiency of the homodyne is 50±5\%. This discrepancy is mainly due to electronic noise, the non-perfect polarization alignment, and temporal overlap between the local oscillator and signal pulses. Homodyne measurement enables the evaluation of a quantum state's projection in phase space with respect to the local oscillator axis. As a result, it is highly sensitive to variations in the relative phase between the desired signal (quantum states with fixed phase and amplitude squeezing) and the local oscillator. Unfortunately, during our proof-of-principle experiment, phase oscillations were unavoidable due to experimental conditions, which also caused oscillations in the squeezing direction. 
To successfully address this challenge, we employed piezoelectric phase shifters, as illustrated in Figure \ref{fig:fig3}. We applied ramp signals to rapidly scan through these phases and subsequently selected measurements corresponding to a null phase and amplitude squeezing. However, it's important to note that our approach has limitations in achieving high communication rates and an active phase stabilization system or a local local oscillator approach is required for a practical implementation of the protocol.
To evaluate the performance of the protocol (as shown in Figure \ref{fig:fig6}), we used for each experimental point 16000 pulses, with an alphabet of four letters for Alice and four different possible initialization values for Bob, distributed uniformly. This corresponds to 1000 pulses (0.1 ms) for each combination of Bob's initialization value and Alice's letter of the alphabet, which allowed us to retrieve the mean value and error. These values would correspond to fixed voltages applied to the VOAs.  
The data supporting the findings of this study are available upon request. Interested researchers may contact the authors directly for access to the data.

\section{Backmatter}


\begin{backmatter}
\bmsection{Funding}
The authors acknowledge the “Quantum FVG” project founded by the Regione Friuli Venezia Giulia and the Project QUID (Quantum Italy Deployment) funded by the European Commission in the Digital Europe Programme under the grant agreement No 101091408.
A.Z. acknowledges the project QuONTENT under the “Progetti di Ricerca@CNR” program funded by the Consiglio Nazionale delle Ricerche (CNR) and the European Union - PON Ricerca e Innovazione 2014-2020 FESR - Project ARS0100734 QUANCOM and.
I.P. and A.Z. acknowledge the project CTEMT (Casa delle Tecnologie Emergenti Matera) founded by the Ministry of Enterprises and Made in Italy (2019). This work was partially supported by projects NQSTI (PE0000023) and SERICS (PE00000014) under the MUR National Recovery and Resilience Plan funded by the European Union – NextGenerationEU. This work was partially supported by PON “Ricerca e Innovazione" 2014-2020 (code: 33-G-15032-2). MGAP acknowledges support from Italian Ministry of Research and Next Generation EU via the PRIN 2022 project RISQUE (contract n. 2022T25TR3). 


\bmsection{Disclosures}
The authors declare no conflicts of interest.

\bmsection{Data availability} Data underlying the results presented in this paper are not publicly available at this time but may be obtained from the authors upon reasonable request.

\bmsection{Supplemental document} See Supplement 1 for supporting content. 

\end{backmatter}







\end{document}